\documentclass[useAMS,usenatbib,usegraphicx]{mn2e}
\usepackage{amssymb}

\title[Extreme magnification]{Primordial star clusters at extreme magnification}
\author[Zackrisson et al.]{Erik Zackrisson$^{1,2}$\thanks{E-mail: erik.zackrisson@physics.uu.se}, Juan Gonz\'alez$^{2}$, Simon Eriksson$^{2}$, Saghar Asadi$^{2}$, \newauthor Chalence Safranek-Shrader$^{3}$, Michele Trenti$^4$ and Akio K. Inoue$^{5,3}$\vspace{0.25cm}\\
$^{1}$Department of Physics and Astronomy, Uppsala University, Box 515, SE-751 20 Uppsala, Sweden\\
$^{2}$Department of Astronomy, Stockholm University, Oscar Klein Center, AlbaNova, Stockholm SE-106 91, Sweden\\
$^{3}$Department of Astronomy and Astrophysics, University of California, 1156 High Street, Santa Cruz, CA 95064, USA\\
$^{4}$School of Physics, The University of Melbourne\\
$^{5}$College of General Education, Osaka Sangyo University, 3-1-1, Nakagaito, Daito, Osaka 574-8530, Japan}

\begin{document}

\date{Accepted ... Received ...; in original form ...}

\pagerange{\pageref{firstpage}--\pageref{lastpage}} \pubyear{2014}

\maketitle

\label{firstpage}

\begin{abstract}
Gravitationally lensed galaxies with magnification $\mu \approx 10$--100 are routinely detected at high redshifts, but magnifications significantly higher than this are hampered by a combination of low probability and large source sizes. Magnifications of $\mu \sim 1000$ may nonetheless be relevant in the case of intrinsically small, high-redshift objects with very high number densities. Here, we explore the prospects of detecting compact ($\lesssim 10$ pc), high-redshift ($z\gtrsim 7$) Population III star clusters at such extreme magnifications in large-area surveys with planned telescopes like Euclid, WFIRST and WISH. We find that the planned WISH 100 deg$^2$ ultradeep survey may be able to detect a small number of such objects, provided that the total stellar mass of these star clusters is $\gtrsim 10^4\ M_\odot$. If candidates for such lensed Population III star clusters are found, follow-up spectroscopy of the surrounding nebula with the James Webb Space Telescope or groundbased Extremely Large Telescopes should be able to confirm the Population III nature of these objects. Multiband photometry of these objects with the James Webb Space Telescope also has the potential to confirm that the stellar initial mass function in these Population III star clusters is top-heavy, as supported by current simulations. 
\end{abstract}


\ \section{Introduction}
\label{intro}
The detection of the first generation of stars, the chemically pristine Population III (hereafter Pop III), remains one of the holy grails of observational cosmology. These objects likely emerged $\approx 50$ Myr after the Big Bang at redshifts $z \gtrsim 60$ in very rare regions \citep{Naoz et al.,Trenti & Stiavelli a}, became more common by $z\approx 30$ \citep[at $\sim 100$ Myr;][]{Bromm & Larson, Trenti & Stiavelli b}, and possibly continued forming in pockets of unenriched gas for several billions of years thereafter \citep[e.g.][]{Tornatore et al.,Johnson10,Fumagalli et al.,Simcoe et al.}. Due to the cooling properties of zero-metallicity gas, the stellar initial mass function (IMF) of Pop III stars is predicted to be top-heavy, peaking at $\sim 10$--$100 \ M_\odot$ \citep[e.g.][]{Greif et al. a,Hosokawa et al.,Stacy et al.,Susa et al.,Hirano et al.}. Putting this to the observational test, however, has turned out to be extremely challenging. Current proposals rely on the chemical enrichment signatures of Pop III stars \citep[for a review, see][]{Karlsson et al.} or on the detection of large numbers of supernovae produced by these objects \citep[][]{Hummel et al.,de Souza et al.}. The prospects of detecting individual Pop III stars during their main-sequence lifetimes remain bleak \citep{Rydberg et al. a}, even when the fluxes of such stars are boosted by gravitational lensing. Larger numbers of Pop III stars could potentially form simultaneously within the HI, or atomic, cooling halos (virial mass $\sim 10^7$--$10^8\ M_\odot$) believed to host some of the first galaxies at $z \lesssim 15$ \citep[e.g.][]{Greif et al.}. While most of these atomic cooling halos will be pre-enriched by previous generations of Pop III stars, some may be chemically pristine \citep[e.g.][]{Stiavelli & Trenti} and form Pop III star clusters\footnote{Also known as Pop III {\it galaxies} -- while their stellar masses are more reminiscent of star clusters, these objects are sitting in their own dark matter halos, unlike star clusters in the local Universe.} \citep[e.g.][]{Johnson et al. a}. These objects may be detectable in gravitationally lensed fields, potentially even with existing telescopes, provided that their star formation efficiencies, or equivalently stellar population masses, are sufficiently high \citep{Zackrisson12,Rydberg et al. b}. 

The integrated luminosities of these objects are likely dominated by nebular emission from photoionized gas \citep[e.g.][]{Schaerer,Inoue,Zackrisson11a}, and while the strength of emission lines like Ly$\alpha$ and HeII$\lambda$1640 in their spectra may provide some information on the Pop III IMF \citep[e.g.][]{Raiter et al.}, both lines are somewhat problematic as IMF probes, even in the case where the lack of metal emission lines indicates a very low metallicity \citep{Inoue,Zackrisson11a}. Gas cooling may boost the fluxes of Ly$\alpha$ and HeII$\lambda$1640 \citep{Yang et al.,Dijkstra}, and Ly$\alpha$ may be absorbed by the partly neutral IGM at $z\gtrsim 6$ \citep[e.g.][]{Hayes et al.}. Here, we consider an alternative approach to probing the Pop III IMF, through the measurement of direct star light from Pop III star clusters. However, the detection of such star clusters (rather than the combined, blended light from both star cluster and nebula), would require extreme magnification ($\mu \gg 100$) due to gravitational lensing by foreground objects (low-redshift galaxies and galaxy clusters), even when next-generation telescopes are considered.

High-redshift galaxies in lensed fields are routinely detected at magnification factors of $\mu\sim 10$--100 \citep[e.g.][]{Gonzalez et al. b,Bussmann et al., Bradley et al.}, whereas magnifications of $\mu \gg 100$ are hampered both by the extremely low probabilities for such sightlines (often assumed to obey $P(\mu)\propto \mu^{-3}$ at the high-magnification end; e.g. \citealt{Pei93b}, or equivalently $P(>\mu)\propto \mu^{-2}$) and the fact that the region in the source plane subject to such magnifications tends to be smaller than kpc-scale galaxies \citep[e.g.][]{Peacock,Perrotta et al.}\footnote{It is, however, interesting to note that many of the faintest $z\gtrsim 6$ galaxies appear to be significantly smaller than this \citep{Kawamata et al.}.}. 

High-redshift Pop III star clusters \citep{Stiavelli & Trenti,Johnson et al. a,Johnson10} of stellar mass $\sim 10^4\ M_\odot$ may, on the other hand, be $\lesssim 10$ pc in size, in analogy with star clusters of similar mass in the local Universe \citep[e.g.][]{Bastian et al.}, and could in principle attain magnifications of $\mu \gg 100$, thereby pushing them above the detection threshold of upcoming telescopes. Still, very large survey areas would be required to find such objects due to the tiny probabilities for sightlines with the required magnification. Here, we explore the prospects of detecting such extremely magnified Pop III star clusters in planned large/deep surveys, with special emphasis on the capabilities of the Wide-field Imaging Surveyor for High-redshift (WISH; \citealt{Yamada et al.}) 100 deg$^2$ Ultra Deep Survey (UDS)\footnote{http://wishmission.org}. 

Once an object of this type has been detected in an imaging survey with large areal coverage, follow-up spectroscopy of the surrounding photoionized nebula will be able to measure the redshift and confirm the absence of metal lines. Deeper imaging of the central Pop III star cluster with the James Webb Space Telescope (JWST) may furthermore be able to measure the colour with sufficient accuracy to confirm the top-heavy IMF hypothesis. 
  
In Sect.~\ref{probabilities}, we use cosmological simulations to estimate the probabilities for extreme magnifications at high redshifts. In Sect.~\ref{WISH}, we derive the properties of Pop III star clusters required to make such objects detectable in the WISH UDS. The prospects of constraining the Pop III IMF in these objects using follow-up observations with JWST are discussed in Sect.~\ref{JWST}. A discussion on potential complications and avenues for future research are outlined in Sect.~\ref{discussion}.
	
\section{The probabilities for extreme magnifications}
\label{probabilities}
To estimate the probability for extreme magnifications ($\mu>100$) at high redshifts, we use cosmological simulations to model the matter distribution (dark matter and galaxies) up to redshift $z=10$. The matter is then projected onto a sequence of lens planes and ray-tracing calculations are used to derive the lensing properties along large numbers of random sightlines.

\subsection{Lensing simulations}
The dark matter distribution along the line-of-sight is based on the Millennium $N$-body simulation \citep{Springel et al.}, which follows the evolution of 10 billion dark matter particles in a cubic volume of side 500 $h^{-1}$ Mpc from $z = 127$ to the present epoch in a flat cold dark matter (CDM) universe ($\Omega_\mathrm{M} = 0.25$, $\Omega_\Lambda = 0.75$, $H_0=73$ km s$^{-1}$ Mpc$^{-1}$. $\sigma_8 = 0.9$). The $N$-body particle data is stored at 64 epochs and many replications of these snapshot boxes are required to simulate the full line-of-sight to light sources at high redshifts. To avoid the repetition of structures within a given line-of-sight, each sightline is set at an angle with respect to the box side. To model the effects of high-density baryonic structures, the dark matter halos of the Millennium simulation are populated with galaxies based on the  Durham semi-analytic model \citep{Baugh et al.}. With a particle mass of $8.6 \times 10^8\ h^{-1}\ M_\odot$, about 20 million haloes with more than 20 particles each are formed throughout the Millennium simulation. The halos are identified using a friends-of-friends group finding algorithm \citep{Davis et al.} and have been decomposed into subhaloes using the recipe by \citet{Helly et al.}. These merger trees \citep{Harker et al.} are then used as input for the semi-analytic model, which assumes that galaxy disks form from gas that accumulates at the centre of each halo/subhalo, whereas bulges are assumed to form through galaxy mergers occurring after the merging of haloes/subhalos in the tree \citep[see][for details]{Gonzalez et al. a}.

Once the distribution of dark matter and galaxies throughout the light cone has been generated, this matter distribution is collapsed onto a sequence of 52 lens planes up to a redshift of $z\approx 10$. The method we used to compute the deflection and magnification produced along each line-of-sight is described in detail by Gonzalez et al. (2014, in prep.). In short, we use a ray-tracing algorithm \citep[e.g.][]{Premadi et al.,Hilbert et al. a} that follows a set of light rays, forming a grid with separation 2.5 comoving kpc, through the lens planes. In each plane, the lensing potential and the deflections produced by the dark matter particles are computed as in \citet{Hilbert et al. a}, whereas the contributions from galaxies are calculated by adopting the De Vaucouleurs and exponential disk lensing models presented in \citet{Keeton} for the bulges and disk components of these objects. While \citet{Hilbert et al. a} did not specifically discuss magnifications $\mu>100$ or source redshifts $z_\mathrm{s}>6$, we have checked that our resulting magnification probability distributions are consistent with theirs at lower reshifts and at $\mu<100$, once the galaxy contributions are switched off in our code. The statistics we present in the following correspond to individual lensed images, since multiple images are not kept track of in our computational machinery. It should be kept in mind, however, that multiple images are in most cases expected to be present (at separations of arcseconds to arcminutes, depending on the mass of the main lens) whenever extreme magnifications are produced, and could potentially help weed out interlopers (see Sect.~\ref{discussion}) 

\subsection{Probabilities}
\begin{figure}
\includegraphics[width=84mm]{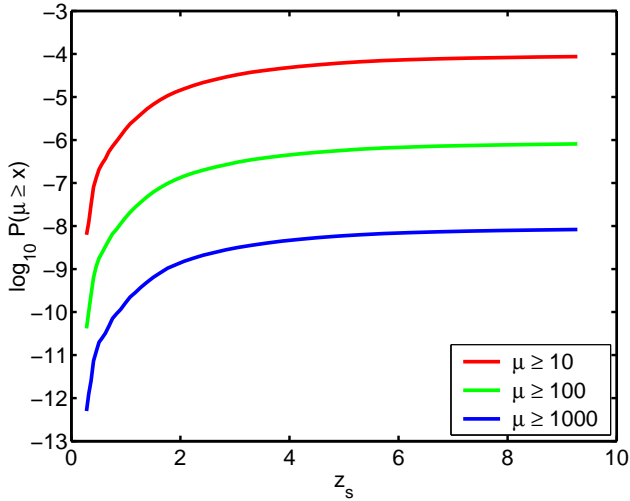}
\caption{The probability for very high magnifications, as a function of source redshift $z_\mathrm{s}$. The probabilities for $P(\mu\geq 10)$ (red line), $P(\mu\geq 100)$ (green line) and $P(\mu\geq 1000)$ (blue line) all refer to the probabilities in the source plane.} 
\label{fig1}
\end{figure}

In Fig.~\ref{fig1}, we plot the source-plane probability $P(\geq \mu)$ as a function of source redshift $z_\mathrm{s}$ to attain magnifications higher than or equal to $\mu$, for $\mu=10$, $\mu=100$, and $\mu=1000$. As expected, $P(\geq \mu)\propto \mu^{-2}$, which means that probabilities for intermediate values for the magnification limits are straightforward to interpolate from these curves within the range of $z_\mathrm{s}$ plotted. 

As seen in Fig.~\ref{fig1}, only one in about $\sim 10^8$ sightlines attains a magnification $\mu \geq 1000$ at $z_\mathrm{s}>6$.  To find even a single source with such an extreme magnification therefore requires the design of a survey wide enough (in terms of angle in the sky) and deep enough (in terms of redshift) to cover an unlensed volume with $\sim 10^8$ objects of the relevant type. Under the approximation that $P(\geq \mu)$ is constant with $z_\mathrm{s}$, the probability $P_\mathrm{obj}$ of having $\geq 1$ object with magnification $\geq \mu$ in a survey of areal coverage $A_\mathrm{survey}$ (deg$^2$) is:
\begin{equation}
P_\mathrm{obj} \approx 1-\left[1 - P(\geq \mu)\right]^{3600A_\mathrm{survey}n_\mathrm{obj}},
\label{detection_probability}
\end{equation}
where $n_\mathrm{obj}$ is the total number of target objects per arcmin$^2$ in an unlensed field in the redshift interval probed (regardless of the luminosities or apparent magnitudes of these objects).  For example, a source population with $n_\mathrm{obj}\approx 300$ objects arcmin$^{-2}$ and a 100 deg$^2$ survey (as considered in Sect.~\ref{WISH}, giving $\approx 10^8$ objects in the whole survey field) results in a $\approx 66\%$ chance of getting one or more such object at $\mu\geq 1000$. The corresponding probability for $\mu\geq 500$ becomes $\approx 93\%$. As shown by \citet{Stiavelli & Trenti} and \citet{Zackrisson12}, areal number densities $n_\mathrm{obj}\sim 100$ arcmin$^{-2}$ match the expectation for metal-free atomic cooling halos (capable of hosting Pop III star clusters) at $z\geq 7$. The approximation of an unevolving $P(\geq \mu)$ at high redshift is a decent one, since the $P(\geq \mu)$ functions of Fig.~\ref{fig1} increase very slowly at high redshifts (with only $\approx 20\%$ relative increase in the range $z_\mathrm{s}=6$--9), primarily due to very modest evolution in the relevant angular-size distance ratios (the observer-to-source to lens-to-source ratio $D_\mathrm{OS}/D_\mathrm{LS}$) at these redshifts.

The calculation presented above is based on the assumption that the survey area is randomly selected, and not -- for instance -- solely comprised of cluster fields preselected to boost the chances of high magnifications \citep{Wong et al. a, Wong et al. b}. This should be a fair approximation in the current context, since wide-field surveys tend to serve a number of diverse science goals, which means that only a fraction of the survey time can realistically be spent on fields with highly unusual lensing properties. We stress, however, that this approach is fundamentally different from that adopted by \citet{Zackrisson12}, who considered the detectability of high-redshift Pop III objects behind a single, optimal lensing cluster (but only magnifications $\mu\lesssim 100$). 

While a detailed investigation into the line-of-sight properties of extreme magnification sightlines is outside the scope of this paper, a casual analysis reveals them to be quite complicated, with substantial magnification contributions from lenses at several different redshifts \citep[for similar discussions at lower magnifications, see e.g.][]{Hilbert et al. a,Hilbert et al. b,Wong et al. a,Wong et al. b,Decker French et al.}. The main lens (the galaxy or galaxy cluster that provides the highest single contribution to the overall magnification) is almost always located at $z\lesssim 3.0$ (following a wide distribution peaking at $z\approx 1$), but the dominance of this object is diminished as the overall source magnification goes up. While 85\% of the sources at $z_\mathrm{s}\approx 9$ that are magnified by a factor $\mu\approx 10$ will get more than half of their magnification from the main lens, this fraction drops to $\approx 50\%$ for sources with $\mu\approx 100$ and further to $\approx 10\%$ for sources with $\mu\approx 1000$.

As indicated by eq.(\ref{detection_probability}), surveys with larger areal coverage than 100 deg$^2$ would allow for objects at even higher magnification to be detected. By using $P(\geq \mu)\propto \mu^{-2}$ to extrapolate to higher $\mu$, we estimate a $\approx 66\%$ chance of getting $\geq 1$ Pop III star cluster at $\mu\geq 10000$ in a 10,000 deg$^2$ survey. This may be relevant for the Wide Field Infrared Survey Telescope\footnote{http://wfirst.gsfc.nasa.gov/} (WFIRST; \citealt{Goullioud et al.}) and Euclid\footnote{sci.esa.int/euclid} (\citealt{Amiaux et al.}), which will survey $\sim 10^4$ deg$^2$ with survey limits of $m_{AB}\approx 24$--26.7 at 1--2 $\mu$m. The Large Synoptic Survey telescope\footnote{http://www.lsst.org} (LSST; \citealt{Abell et al.}) performs similarly in terms of survey area and depth, but at wavelengths $\lesssim 1.1$ $\mu$m, which limits the redshift range to $z\lesssim 7$. However, magnifications $\mu \gg 1000$ would also require much smaller intrinsic source sizes. Since high magnifications map to very small source-plane areas, there is effectively a maximum magnification $\mu_\mathrm{max}$ for sources of any given size. Casual analytical estimates place this at $\mu_\mathrm{max}\approx 1000$ for a 10 pc source at $z\approx 10$ \citep{Peacock}, in rough agreement with estimates based on our own simulations. If one choses to push for magnifications $\mu\gg 1000$, the exact size (and shape) of source place regions subject to such  extreme magnifications would become crucial. Here, we will make no attempts to model any effective cut-off in the magnification distribution, and instead focus on the WISH 100 deg$^2$ UDS, for which the $\mu\gg 1000$ regime is out of bounds (see Sect.~\ref{WISH}).

\section{Detecting Pop III star clusters with WISH}
\label{WISH}
The planned WISH UDS will cover 100 deg$^2$ down to an $5\sigma$ AB magnitude limit of $\approx 28$ in 5 broadband filters at 1--3 $\mu$m. Even though this will in principle allow the drop-out selection of objects up to $z\approx 15$, we here consider only the redshift range $z=7$--13, since this is where the areal number densities (objects per arcmin$^2$ and unit redshift in unlensed fields) of metal-free atomic cooling halos peak (\citealt{Zackrisson12}, based on simulations by \citealt{Trenti et al.}). If we assume that Pop III star clusters form in an instantaneous burst, and consider stellar population ages up to 5 Myr (after which the object may be difficult to separate from more metal-enriched counterparts, as discussed in Sect.~\ref{colour}), the predictions presented in \citet{Zackrisson12} indicates that one could expect up to $\approx 400$ such objects per arcmin$^2$ at $z=7$--13 in an unlensed field. This converts into $\approx 1.4\times 10^8$ such objects in the whole survey area covered by the WISH UDS. Thanks to the huge number of Pop III star clusters, we find a 90\% probability of having $\geq 1$ such object with magnification $\mu \gtrsim 700$ in the UDS (eq.~\ref{detection_probability}). While magnifications $\mu \gg 700$ are in principle possible within the survey area, they are considerably less likely, as $P_\mathrm{obj}$ drops off quickly with increasing $\mu$. For example, the probability of having at least one magnified Pop III star cluster with $\mu \gtrsim 1000$ is $\approx 70\%$, and the probability of $\mu \gtrsim 3000$ is $\approx 10\%$. In the following, we adopt therefore the $\mu \approx 700$ as the highest magnification likely to be relevant for Pop III star clusters in the WISH UDS. 
\begin{figure}
\includegraphics[width=84mm]{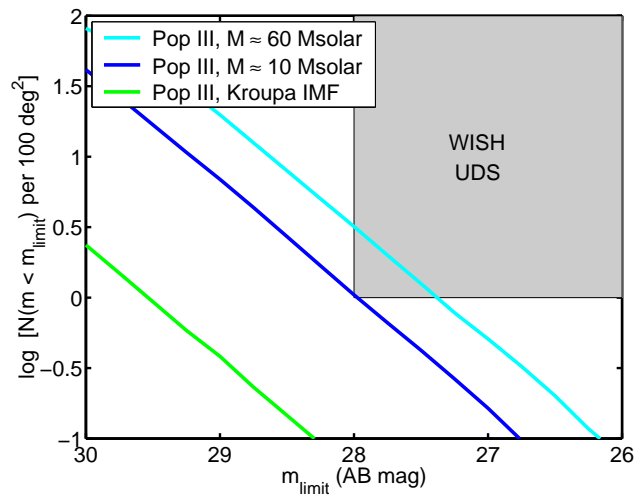}
\caption{The predicted number of extremely magnified Pop III star clusters at $z=7$--13 in the WISH 100 deg$^2$ UDS, as a function of their brightest AB magnitude in the WISH passbands. The three lines correspond to different IMFs: The \citet{Kroupa} IMF characteristic of metal-enriched star formation in the local Universe (green), a top-heavy IMF peaking at $10\ M_\odot$ (blue) and a top-heavy IMF peaking at $60 \ M_\odot$ (cyan). In all cases, we assume that Pop III star clusters for single-age stellar populations with typical masses of $1.5\times 10^4\ M_\odot$ and consider only clusters with ages up to 5 Myr. The shaded region indicates the region in which $\geq 1$ Pop III star cluster would be detectable in the WISH UDS. All objects falling in this part of the diagram have magnification $\mu>300$.} 
\label{fig2}
\end{figure}

By convolving the Pop III spectral energy distributions of \citet{Raiter et al.} with the WISH passband fluxes, we find that, at a magnification of $\mu \approx 700$, the direct star light from a 2 Myr old Pop III star cluster at $z\approx 10$ can be detected in the WISH UDS (in at least one filter) provided that the stellar mass within such a cluster is $\geq 2\times 10^4\ M_\odot$ if the Pop III IMF peaks at $\approx 10\ M_\odot$ (lognormal function with $\sigma=1.0$ within the interval 1--500 $M_\odot$; TA in the notation of Raiter et al.) and $\geq 8\times 10^3\ M_\odot$ if the Pop III IMF peaks at $\approx 60\ M_\odot$ (TE in Raiter et al.). This brackets the likely range of IMFs in these objects \citep[e.g.][]{Hosokawa et al.,Susa et al.,Hirano et al.}. By comparison, a Pop III star cluster forming with an IMF characteristic of metal-enriched stars in the low-redshift Universe \citep{Kroupa}, would require a total stellar mass of $\gtrsim 1\times 10^5\ M_\odot$ to be detected at this magnification. 

Detailed Monte Carlo simulations that take into account the age, redshift, and magnification distributions of the Pop III star cluster population result in slightly different detection limits, but the stellar population mass still needs to be $\gtrsim 10^4\ M_\odot$ for the two top-heavy IMFs considered, as long as the same stellar population mass is adopted for all star clusters.

This is illustrated in Fig.~\ref{fig2}, where we present a Monte Carlo simulation of the expected number of Pop III star clusters at $z=7$--13 as a function of their brightest AB magnitude in the WISH 100 deg$^2$ UDS, assuming a total stellar mass of $1.5\times 10^4\ M_\odot$ for these clusters (matching the most massive Pop III star clusters considered in the simulations by \citealt{Johnson et al. a}).  Here, metal-free atomic cooling halos of virial mass $3\times 10^7$ -- $1\times 10^8\ M_\odot$ from the numerical simulations presented in \citet{Zackrisson12} are randomly assigned sightlines based on our lensing simulations (Sect.~\ref{probabilities}) and ages in the range 0--5 Myr. In this case, the models based on the two top-heavy Pop III IMFs (blue and cyan lines) predict that $\approx 1$--3 objects of mass $1.5\times 10^4\ M_\odot$ would turn up at detectable flux levels in the WISH UDS, whereas the \citet{Kroupa} IMF (green line) would place such objects far below the detection threshold.  

While the intrinsic redshift distribution of metal-free atomic cooling halos peaks at $z\approx 10$ \citep{Zackrisson12}, the redshift distribution of objects that end up above the UDS detection limit in this Monte Carlo simulation is dominated by star clusters at $z\approx 7$--8, with an average halo mass of $\approx 6\times 10^7\ M_\odot$. A typical Pop III star cluster mass of $\approx 1.5 \times 10^4\ M_\odot$ therefore requires that $\approx 0.2\%$ of the total baryonic mass of these halos is converted into Pop III stars over a time scale of just a few million years. This is marginally consistent with the most optimistic case motivated by the simulations of \citet{Safranek-Shrader et al.}, but potentially within range of the gas collapse scenarios discussed by \citet{Johnson et al. 14}. 

When comparing observational detection limits to simulations, it is admittedly not only the {\it typical} Pop III star cluster mass that matters, but also the mass distribution. In principle, the typical Pop III star cluster mass could be $\ll 10^4\ M_\odot$ while still allowing Pop III star clusters to turn up in the WISH UDS (i.e. within the gray region indicated in Fig.~\ref{fig2}), provided that the cluster mass function has a non-negligible high-mass tail that extends to $\gg 10^4\ M_\odot$. In this case, the anomalously high intrinsic fluxes of the most massive objects may push them above the survey flux limit even if they are not located along sightlines with the most extreme magnifications. Consider, for example, a scenario in which 90\% of all Pop III star clusters have masses $\sim 10^3\ M_\odot$ but 10\% have $\sim 10^5\ M_\odot$. The $10^5\ M_\odot$ clusters would be detectable along sightlines with magnification $\mu\approx 100$, which are $\approx 50$ times more numerous than the more extreme $\mu\approx 700$ sightlines required for the detection of $\approx 1.5\times 10^4\ M_\odot$ objects. In this case, the larger number of $\mu\approx 100$ sightlines more than compensates for the small fraction of $\sim 10^5\ M_\odot$ clusters available. In summary, the detection of direct star light from Pop III star clusters in the WISH UDS still requires the formation of some $\gtrsim 10^4\ M_\odot$ objects, but the average masses can in principle be lower.

\section{Probing the Population III stellar initial mass function}
\label{JWST}
\begin{figure}
\includegraphics[width=84mm]{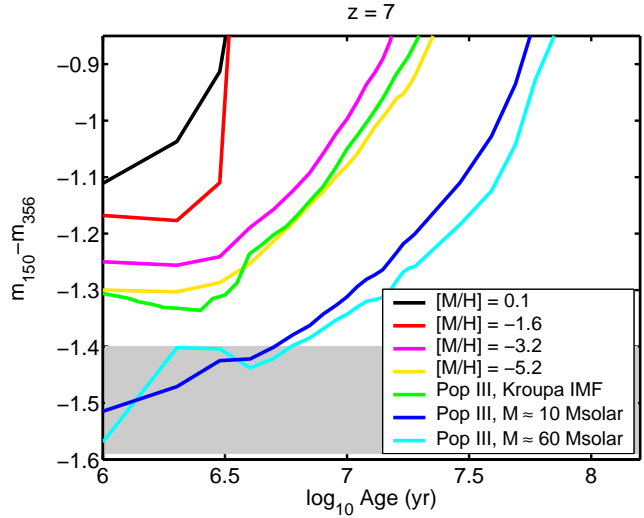}
\caption{Confirming a top-heavy Pop III IMF using JWST/NIRcam imaging. The differently coloured lines represent the predicted evolution of $z=7$, instantaneous burst stellar populations at various metallicites. Black, red, magenta, yellow and green tracks indicate models with metallicities ranging from solar to zero, all obeying a \citet{Kroupa} IMF, characteristic of metal-enriched star formation in the local Universe, The remaining two tracks assumes a Pop III stellar population with a top-heavy IMF peaking at $10\ M_\odot$ (blue) and $10\ M_\odot$ (cyan). At this redshift, a measured colour of $m_{150}-m_{356}<-1.2$ would indicate an extremely metal-poor or Pop III stellar population ($\left[M/H\right]\lesssim -3$), and $m_{150}-m_{356}<-1.4$ (gray region) would indicate a Pop III stellar population with a top-heavy IMF.} 
\label{fig3}
\end{figure}

Once a candidate Pop III star cluster has been identified in a wide-area survey like the WISH UDS, follow-up observations with other telescopes should be able to shed further light on its properties. Spectroscopy of the surrounding nebula (see Sect.~\ref{discussion}) using either groundbased Extremely Large Telesccopes (ELTs) or the JWST will be able to pin down the exact redshift and demonstrate the lack of metal lines, whereas deeper imaging carried out with the JWST towards the central star cluster may be able to confirm that the Pop III IMF really is top-heavy. This is illustrated in Fig.~\ref{fig3}, where we plot the temporal evolution of single-age stellar populations for various metallicities and IMFs in the $m_{150}-m_{356}$ JWST/NIRCam colour  at $z=7$ (effectively probing the slope of the UV/optical stellar continuum). At ages $\leq 5$ Myr, Pop III ($Z=0$) and extremely metal-poor stellar populations ($Z=10^{-7}$ to $10^{-5}$, i.e. $[M/H]\approx -5.2$ and $-3.2$ if $Z_\odot=0.016$) attain far bluer colours than Pop II/I ($Z\geq 0.0004$, i.e. $[M/H]\geq -1.6$). Adopting a top-heavy Pop III IMF shifts the colour to even more extreme values (blue and cyan lines). Hence, the detection of a star cluster with an anomalously blue colour at high redshift would clearly indicate a top-heavy IMF. At $z\approx 7$, the Pop III, top-heavy IMF regime would be at $m_{150}-m_{356}\leq -1.4$ (see Fig.~\ref{fig3}), but similar limits can be set at other redshifts and with other filter combinations as well. By itself, WISH would provide a rough handle on the Pop III star cluster colour, but the error will be much too large to say something definitive about the IMF unless the object is much brighter than the $5\sigma$ detection threshold. JWST, on the other hand, can push the colour uncertainty down to $\Delta(m_{150}-m_{356})\leq 0.1$ mag for an $m_\mathrm{150}\lesssim 27.5$ mag object of colour $\Delta(m_{150}-m_{356})\approx -1.5$ (i.e. indicating a Pop III star cluster with a top-heavy IMF at $z\approx 7$) in a total exposure time of $\lesssim 10$ hours\footnote{Based on prototype JWST exposure type calculator, v.1.6, available at http://jwstetc.stsci.edu/etc/} using the NIRCam instrument. 

\label{colour}

\section{Discussion}
\label{discussion}

\subsection{Expected morphology}
\label{morphology}
The Pop III star cluster is likely surrounded by a prominent, photoionized nebula of size $\sim 100$ pc \citep{Johnson et al. a}. The exact luminosity of this nebula critically depends on the fraction of ionizing photons escaping into the intergalactic medium, and although this fraction may become very high due to feedback effects \citep[50--95\%;][]{Johnson et al. a}, the nebula is predicted to dominate or at least substantially contribute to the integrated rest-frame ultraviolet/optical fluxes of the object \citep{Zackrisson11a,Zackrisson13}. Since the nebular spectrum tends to be very red \citep[see][for a more thorough discussion]{Zackrisson11a}, any contamination from nebular emission to the star cluster measurement will jeopardize colour diagnostics like that shown in Fig.~\ref{fig3}. The effect is to make extremely metal-poor or Pop III star clusters appear more metal-enriched than they really are, and to make star clusters with top-heavy IMFs appear more like those obeying the \citet{Kroupa} IMF (relevant in the present-day Universe). In the absence of gravitational lensing, there is no hope of spatially separating the Pop III star cluster and the surrounding nebula through WISH/JWST imaging, since both components will be below the $\sim 0.1\arcsec$ resolution limits of these telescopes at $z\gtrsim 7$.  However, observations of Pop III star clusters subject to extreme gravitational magnification should allow the two structures to be spatially separated, thereby allowing a cleaner colour measurement towards the stellar component. 

The likely morphology of such an extremely magnified system is that of an extended red arc formed by the nebula, with the Pop III star cluster superposed as a very blue, point-like source. Due to its more extended size, the nebula will attain a lower magnification than the star cluster, but since its intrinsic luminosity is likely to be higher, the two may actually attain similar total fluxes. This makes the nebula a promising target for follow-up spectroscopy with JWST or groundbased ELTs to confirm the redshift (using the HeII$\lambda$1640 \AA{} or H$\beta$ emission lines), demonstrate the absence of metal emission lines like [OIII]$\lambda$5007 \citep{Inoue,Zackrisson11a}, and use the strength of the HeII$\lambda$1640 \AA{} line to probe the nature of the stellar population providing the ionizing photons \citep[e.g.][]{Schaerer,Jimenez & Haiman,Raiter et al.,Pawlik et al.,Inoue}. 

A $\sim 100$ pc nebula at $z\gtrsim 7$ will be extended beyond the point spread function (PSF) of WISH/JWST even at a magnification of $\mu\approx 10$--100. The $\lesssim 10$ pc star cluster may, on the other hand, remain unresolved even at a magnification of up to $\mu\approx 1000$, depending on the exact intrinsic size, the image stretching factor and the PSF. Our lensing simulations indicate that the typical image stretching (longest to shortest length ratio) may be a few times $\sim 100$ for a source magnified by $\mu\approx 1000$. Adopting a stretching factor of 300 would then imply an angular source size of 0.08\arcsec along the most extended direction for a 1 pc source at $z\approx 10$, which makes it marginally resolved in JWST/NIRCam imaging (PSF full width half maximum 0.07\arcsec at 2$\mu$m). In the case of WISH imaging, the source may attain an intrisic size a few times higher while still remaining point-like (PSF full width half maximum 0.3\arcsec at 2$\mu$m). The exact contribution from nebular emission across this central object will depend on the surface brightness distribution of the nebula (in turn depending on its spatial density distribution of photoionzed gas and the escape fraction of ionizing photons), the telescope PSF and on the exact magnification distribution across the system. More detailed lensing simulations would be required to study this in detail. If it turns out that the nebular contribution across the central star cluster will typically be negligible (or can be corrected for), then even a null result in a survey like the WISH UDS would provide important information on the star-forming properties of atomic cooling halos. If no very blue, extremely magnified objects are detected, this would indicate that fewer atomic halos than predicted are metal-free, or that the the star formation efficiency in such systems places them below the photometric detection threshold of the survey.

\subsection{Identifying Pop III star clusters in large surveys}
How would these extremely magnified Pop III star clusters be identified among the huge number of more mundane objects in an automated catalog based on a deep, wide-area survey like the WISH UDS? By combining drop-out criteria (indicating high redshift) with requirements for substantial elongation (as expected for a gravitational arc), one is still likely to be faced with large numbers of interlopers in the form of edge-on disk galaxies, ``chain galaxies'', and blended objects. However, many of these contaminants can be weeded out by either requiring the presence of other highly elongated objects within a few arcminutes (as expected along the critical curve of a galaxy cluster lens) or by demanding high curvature of the elongated structure (as expected when the lens is a single galaxy).  By comparing object colours derived in differently-sized apertures, objects with strong colour gradients (as expected for the blue core on extended red nebula discussed in Sect.~\ref{morphology}) can be singled out as promising candidates. Among these, the most interesting objects should be the ones with the bluest colours (Fig.~\ref{fig3}). Despite the unusual morphology expected for these extremely magnified Pop III objects, rare interlopers that superficially resemble these may however still turn up in large surveys. A cool foreground star that happens to be superposed on the gravitationally lensed arc of a high-redshift galaxy could mimic the morphology predicted above, and -- due to the presence of strong absorption bands in the stellar atmosphere -- potentially also exhibit blue colours if studied in only a small number of filters. However, such interlopers can be weeded out by follow-up spectroscopy of the arc (which is unlikely to display a metal-free emission-line spectrum in this case), or by the detection of multiple images of the arc (since foreground stars are unlikely to be superposed on all of them). 

In the case where the star cluster itself remains unresolved (due to low stretching) and the extended nebula remains undetected due to low surface brightness, colour criteria would have to be applied directly to the dropout catalog, but unless multiple images of the same object can be detected, it may be difficult to establish if the target is indeed gravitationally lensed.    

Since extreme magnification sightlines tend to be affected by lensing at several different redshifts (Sect.~\ref{probabilities}), there is also a non-negligible risk that the strongly lensed image from a Pop III object may be blended with light from foreground lens galaxies, therefore making identification and analysis more difficult. Our recipe for populating halos along the line-of-sight with galaxies from semi-analytical models (which predict both luminosities and sizes) in principle allows us to investigate this issue, and we aim to revisit this in future work.

\subsection{Simulations of star formation in metal-free atomic cooling halos}
The required stellar population mass of $\gtrsim 10^4\ M_\odot$ represents a hard limit for the detection of direct star light from $z\gtrsim 7$ Pop III star clusters at magnification $\mu\sim 1000$ in the foreseeable future. As discussed in Sect.~\ref{intro}, large numbers of Pop III stars can form in atomic cooling halos that manage to remain chemically pristine, a scenario that likely requires a background Lyman-Werner (LW) radiation field that suppresses molecular hydrogen ($\mathrm{H}_2$) formation and, consequently, Pop III star formation in minihaloes. This scenario has been explored predominantly in the context of supermassive star and direct collapse black hole \citep{Bromm & Loeb,Haiman} formation in regions of the Universe with a LW radiation field high enough above the cosmic mean to entirely suppress $\mathrm{H}_2$ formation. It it thought the onset of efficient $\mathrm{H}_2$ cooling is required for star cluster, rather than supermassive star, formation \citep{Regan09}, requiring a UV background intensity below $J_{21}\sim10^3$ for $\mathrm{H}_2$ to form in sufficient abundance but above $J_{21}\sim 10$ to suppress star formation in minihalos (here, $J_{21}$ denotes the radiation intensity at the Lyman-limit in units of $10^{-21} \, {\rm erg}\, {\rm s}^{-1}\, {\rm cm}^{-1}\, {\rm Hz}^{-1}\, {\rm sr}^{-1}$). Explicitly focusing on Pop III star cluster formation, \citet{Safranek-Shrader et al.} argued that total stellar masses of $10^4\ M_\odot$ may be achievable given a low enough LW background intensity, though unaccounted for internal LW feedback may be significant in reducing the star formation efficiency. The inclusion of ionizing radiation further increases the propensity for Pop III galaxies with total stellar masses $\gtrsim10^4\ M_\odot$ \citep{Johnson et al. 14}. Related studies \citep{Regan09,Shang et al.,Regan14,Latif et al.,Visbal et al.} have routinely found central gravitationally unstable clumps with masses in excess of $10^{4-5}\ M_\odot$ and central accretion rates of 0.01 -- 1 $M_\odot$ yr$^{-1}$ to form in chemically pristine, high-redshift atomic cooling haloes. The results of these studies additionally suggest that fragmentation, if it does occur, happens on small scales ($\lesssim 1$--10 pc) favoring the formation of a compact Pop III stellar cluster. While the limit of $\gtrsim 10^4\ M_\odot$ adopted in this work seems marginally achievable for the total stellar mass of a Pop III galaxy, additional cosmological simulations are needed that focus on the possibility and characteristics of a metal-free stellar cluster forming in an atomic cooling halo.

\subsection{Other surveys}
As discussed in Sect.~\ref{probabilities}, $\mu\gg 1000$ scenarios may be relevant in the case of telescopes like Euclid and WFIRST, that are projected to survey even larger regions of the sky than WISH. For instance, the WFIRST-AFTA mission concept could  survey 1600 deg$^2$ down to $m_\mathrm{AB}\approx 26.7$-26.9 per year at 1.0-1.8 $\mu$m. The planned WFIRST High Latitude Survey \citep{Gehrels & Spergel}, scheduled to cover 2400 deg$^2$, would be able to take advantadge of magnifications up to $\mu \approx 3000$ and detect Pop III star clusters with masses $\gtrsim 6\times 10^3\ M_\odot$. More detailed lensing and Pop III star formation simulations would be required to assess the viability of such scenarios, given the star cluster compactness ($\lesssim 3$ pc; \citealt{Peacock}) likely required. Moreover, since WFIRST is currently considered for launch in 2024, it is unclear if it will be able to take advantage of JWST for follow-up studies.  

\section{Summary}
We propose that compact ($\lesssim 10$ pc) high-redshift ($z\gtrsim 7$) Population III star clusters with top-heavy IMFs and total stellar mass $\gtrsim 10^4\ M_\odot$ may be detectable at extreme magnifications ($\mu \gg 100$) in the planned WISH 100 deg$^2$ ultradeep survey. Improved simulations of star formation in metal-free atomic cooling halos would be very useful to assess the likeliness of such star clusters forming. We also argue that once candidates for such objects are found, follow-up observations with the JWST and/or groundbased Extremely Large Telescopes should be able to confirm the Population III nature of these objects and potentially confirm that the stellar initial mass function of Pop III stars is top-heavy.

\section{acknowledgements}
E.Z. acknowledges research funding from the Swedish Research Council (project 2011-5349), the Wenner-Gren Foundations and the Swedish National Space Board. J.G. acknowledges research funding from the Wenner-Gren Foundations. A.K.I. is supported by JSPS KAKENHI Grant Number 26287034. Ikuru Iwata is acknowledged for useful input on the detection limit of the WISH UDS. We kindly thank the anonymous referee of this work for useful comments.

\end{document}